\documentclass[aps,prab,groupedaddress,twocolumn,secnumarabic]{revtex4}
\usepackage{amsmath,amssymb,amsfonts}
\usepackage{graphicx}
\usepackage{hyperref}

\begin{document}
\title{
Multipactor suppression in dielectric-assist accelerating structures via diamond-like carbon coatings
}
\author{Shingo Mori}%
\email{smori@post.kek.jp}
\affiliation{KEK Accelerator department, Tsukuba, Ibaraki 305-0801, Japan}
\author{Mitsuhiro Yoshida}%
\email{mitsuhiro.yoshida@kek.jp}
\affiliation{KEK Accelerator department, Tsukuba, Ibaraki 305-0801, Japan}
\author{Daisuke Satoh}%
\email{dai-satou@aist.go.jp}
\affiliation{Radiation Imaging Measurement Group
 Research Institute for Measurement and Analytical Instrumentation,
   National Metrology Institute of Japan,
   National Institute of Advanced Industrial Science and Technology (AIST)}
\date{\today}%
\begin{abstract}
  Multipactor discharges are widely observed in the accelerator field, and their suppression has been studied to improve accelerator performance.
  A dielectric-assist accelerating (DAA) cavity is a standing-wave accelerating cavity that attains a Q-value of over 100,000 at room temperature by using the reflection of the dielectric layer; the DAA cavity is expected to be used with a small RF power source and high-duty operation.
  Thus far, the maximum accelerating field of DAA cavities has been limited to a few MV/m by the multipactor.
  By applying a diamond-like carbon (DLC) coating to reduce the secondary electron emission coefficient without sacrificing the Q-value of the cavity, we have demonstrated that a $6~[{\rm \mu s}]$ RF pulse can be injected into a DAA cavity with a field of more than 10~[MV/m] while suppressing the multipactor.
\end{abstract}
\maketitle

Multipactors are a widespread phenomenon in the accelerator field and have appeared in the development of RF windows, accelerator tubes, and beam pipes.
In the multipactor phenomenon, the number of electrons increases exponentially owing to repeated resonant collisions of electrons on the surface of a material in an alternating electric field; it occurs when the secondary electron yield (SEY) of the material is greater than one.
There are the several classes of multipactor suppression methods that aim to
reduce the SEY through application of a coating of low SEY material such as TiN~\cite{nyaiesh1986properties,kennedy1997tin} or amorphous carbon~\cite{vallgren2011amorphous,vallgren2010amorphous,shaposhnikova2009experimental}, or
by grooving~\cite{baglin1998photoelectron},
reducing the tangential electric field on the surface,
or applying a solenoidal magnetic field~\cite{kulikov2001electron,jing2016complete}.

\begin{figure*}[t]
  \centering\includegraphics[width=0.9\hsize]{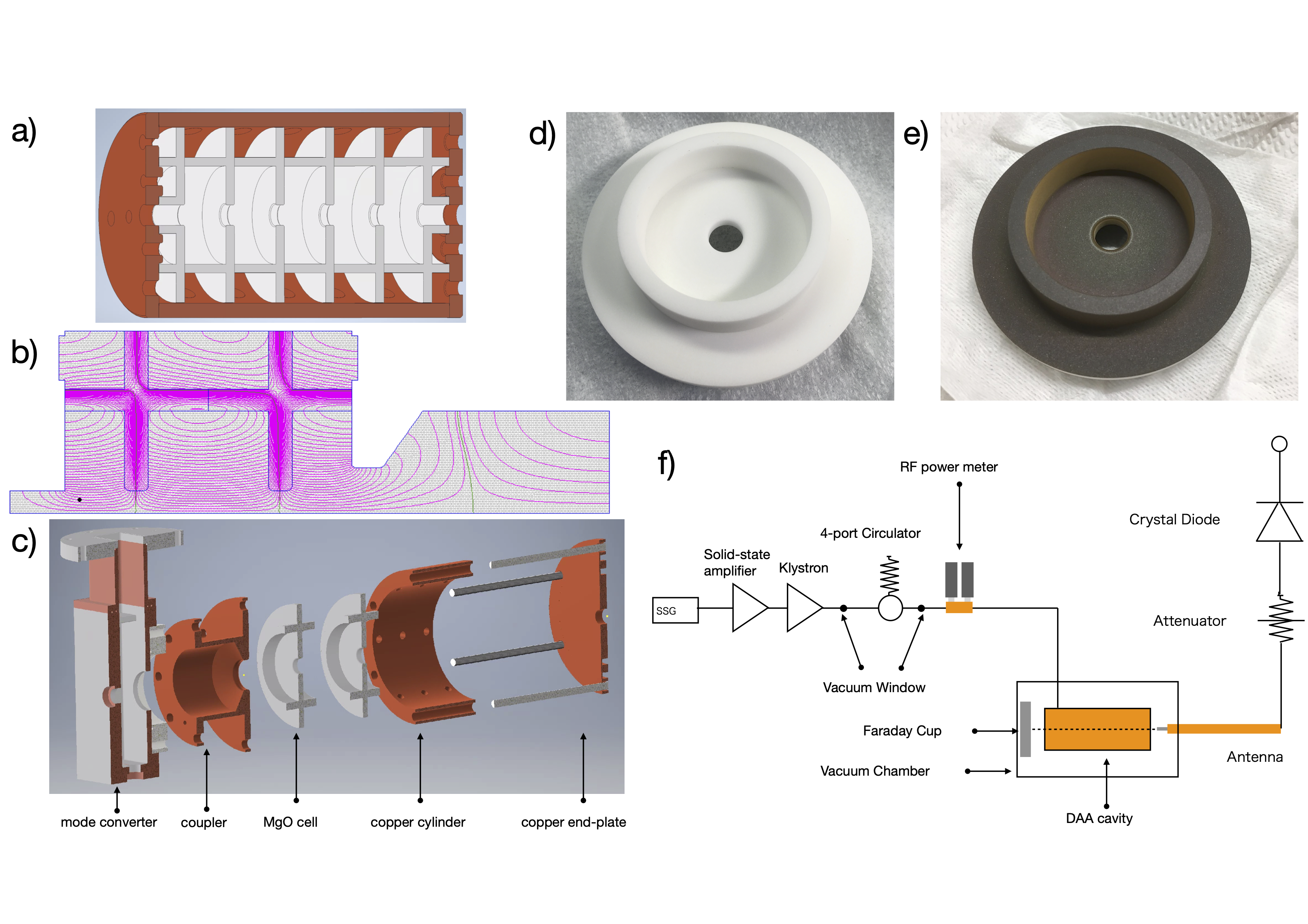}
  \caption{(a) Schematic representation of the original DAA cavity~\cite{satoh2017fabrication}. (b) Cross-section of the electric field of the accelerating mode calculated using {\tt SuperFish}. (c) Development view of the two-cell cavity.
  (d) MgO cell without coating and (e) with DLC coating with a thickness of $0.5~[{\rm \mu m}]$.
  (f) RF system of the high-power test from the klystron to the measurement of stored power through the antenna.}
  \label{fig:cavity_rfsystem}
\end{figure*}

The RF accelerating cavity is widely used to generate high-energy beams in numerous scientific and technological applications.
One of the figures of merit of an RF cavity is the Q-value, which represents the power efficiency of the cavity and is determined according to the conductivity of the material and the structure.
The superconducting cavity has a high Q-value of greater than order ${\cal O}(10^{10})$ at low temperature $T<4.2~[{\rm K}]$, while the normal conducting cavity has a Q-value of ${\cal O}(10^4)$ working at room temperature, as determined by the ohmic loss of the metal.
Recently, it was proven that a dielectric-assist accelerating (DAA) structure could attain a high Q-value of ${\cal O}(10^5)$ at room temperature~\cite{satoh2016dielectric,satoh2017fabrication} through a reduction in the surface magnetic field using the reflection of a dielectric layer exceeding the limits of the Q-value determined by the ohmic loss.
Figure~\ref{fig:cavity_rfsystem}~(a) shows a five-cell DAA cavity with a $Q\sim 1.2\times 10^5$, which is one order of magnitude larger than the typical Q-value of a normal conducting cavity.

The maximum accelerating field of a dielectric-based accelerating cavity is limited by multipactors and breakdowns.~\cite{power2004observation}
Numerous high-power tests of X-band dielectric-loaded accelerating structures (DLA) have been performed, and accelerating fields of up to 20 MV/m have been achieved by applying an axial magnetic field to suppress multipactors.~\cite{jing2016complete}
The study of photonic bandgap cavities using sapphire rods shows a maximum electric field of 19.1 MV/m, which is limited by breakdowns in triple junctions.~\cite{zhang2016high}
In both examples, short RF pulses of $100~[{\rm ns}]$ to $500~[{\rm ns}]$ in the X-band are fed in.
In contrast, in a DAA cavity, relatively long RF pulses of several $\mu s$ are fed in at a lower frequency, the C-band.
In the case of our DAA cavity, the maximum value of the accelerating field was limited to approximately 1 or 2 MV/m without the SEY countermeasure.
Attempts have been made to suppress the multipactor effect in DLA structures by lowering the SEY with TiN coating. However, in the case of DLA with TiN coating, it is not possible to completely suppress the multipactor.~\cite{jing2010progress}
Recently, it has been shown that a diamond-like carbon (DLC) coating is relatively more resistant to multipacting than TiN as a coating for the inner surface of a metal standing-wave accelerating cavity in the X-band.~\cite{xu2019measurement}.

The ideal coating for a DAA cavity will be a material that has both very low electrical conductivity and SEY.
A coating of a material with high electrical conductivity, such as the TiN coating, reduces the Q-value of the DAA cavity by increasing the ohmic loss on the surface of the dielectric cells.
In fact, as shown in Table~\ref{tab:q0-5cell}, when a TiN coating with a   thickness of $10~[{\rm nm}]$ was applied to both sides of the ceramic cells of the five-cell DAA cavity fabricated in our previous work~\cite{satoh2017fabrication}, the reduction in the Q-value was approximately $40$~[\%].
A TiN coating of a few nanometers in thickness will not suffice because it will be displaced by the electrical discharge occurring during conditioning.~\cite{jing2010progress}

DLC has a low SEY and low electrical conductivity,
and is composed of ${\rm sp}^2$ (graphene-like) and ${\rm sp}^3$ (diamond-like) bounds of carbons.
The SEY of DLC coating is less than 1.5.~\cite{zhang2019research}
We also measured the SEY of the DLC coating on an MgO sample with a thickness of $0.5~[{\rm \mu m}]$ using an SEM-based device at KEK~\cite{yamamoto2019recent}.
Then, in the incident energy range between $0.5~[{\rm keV}]$ and $2.0~[{\rm keV}]$, the measured SEY monotonically decreased from $1.4$ to $0.7$,
and the SEY was $\delta<1$ below the incident energy of $1.1~[{\rm keV}]$;
this was lower than the measured SEY of the TiN coating on the MgO.
Moreover, the DLC coating on the surface of a dielectric cell does not affect the Q-value because the DLC coating has a high electrical resistance at room temperature.~\cite{meyerson1980electrical}
As shown in Table~\ref{tab:q0-5cell}, the DLC coating of $0.5~[{\rm \mu m}]$ thickness on both sides of the ceramic cells of the five-cell DAA cavity resulted in a Q-value change of approximately $3~[\%]$, which is within the error of low-power measurements.

\begin{table}
  \begin{tabular}{cccc}
    \hline
    Coating & $f_0~[{\rm MHz}]$ & $\beta$ & $Q_0$\\ \hline
    w/o coating (set 1) & $5708.29$ & $1.4$ &  $112000$\\
    TiN (set 1) & $5713.01$ & $0.79$ & $64000$\\\hline
    w/o coating (set 2) & $5717.10$ & $0.93$ & $113000$\\
    DLC (set 2) & $5717.07$ & $1.0$ & $116000$\\ \hline
  \end{tabular}
  \caption{
  Effect of TiN coating and DLC coating on the Q-value of the five-cell DAA cavity.
  There are two sets of dielectric cells in the five-cell DAA cavity, each with different dimensional errors.
  TiN coating was applied to both sides of all cells in set 1, with a thickness of $10~[{\rm nm}]$.
  DLC coating with a thickness of $0.5~[{\rm \mu m}]$ was applied on both sides of all cells in set 2.
  The Q-value was measured via the coupler and mode converter shown in Figure~\ref{fig:cavity_rfsystem}~(c) from $S_{11}$ using the Agilent N5230A network analyzer.
  }
  \label{tab:q0-5cell}
\end{table}

Figure.~\ref{fig:cavity_rfsystem}~(e) shows the MgO cell after application of the DLC coating.
We applied DLC coating processed by the plasma CVD method with a thickness of $0.5~[{\rm \mu m}]$,
hydrogen content of $5\sim50~[\%]$, and ${\rm sp^3/(sp^2+sp^3)}$ ratio of $20\sim50~[\%]$.

In this study, we performed a high-power test of the two-cell DAA cavity with low-SEY and low-conductivity DLC coating.
The various conditions are described as follows.

To focus on the effect of the surface treatment on the accelerating gradient, we fabricated a DAA test cavity with two cells.
In order to implement the translational symmetric field configuration shown in Figure~\ref{fig:cavity_rfsystem}~(b) and reduce the tangential electric field on the surface of the DAA cells, we did not adopt the end cell structure to mitigate the power loss at the copper endplate, as discussed in ~\cite{satoh2016dielectric}.
The Q-value of our fabricated test cavity was $Q_0\sim2.7\times 10^4$, which is approximately $25~[\%]$ lower than the Q-values of the simulations assuming electrical conductivity $\rho=1.72\times10^{-8}~[{\rm \Omega/m}]$ and dielectric loss $\tan\delta=6\times10^{-6}$.
This is due to the insufficient contact and surface treatment of the metal cavity and end-plates.

As shown in Figure~\ref{fig:cavity_rfsystem}~(c), the test cavity is composed of
a copper end-plate,
copper cylinder,
coupler, the mode converter,
and dielectric cell.
Figure~\ref{fig:cavity_rfsystem}~(d) shows the MgO cell after machining with tolerances from $0.01~[{\rm mm}]$ to $0.07~[{\rm mm}]$.
The MgO ceramics had the same characteristics as those used in the previous work~\cite{satoh2017fabrication},
i.e., a relative permittivity of $\epsilon_r=9.64$ and a loss tangent of $\tan\delta=6\times10^{-6}$.

Figure~\ref{fig:cavity_rfsystem}~(f) is an RF schematic showing the power source, waveguide, acceleration cavity, and measurement system that comprise the C-band test stand.
The C-band RF was produced by a Tektronix TSG 4106A signal generator (SG) and amplified by a solid-state amplifier and the klystron.
The DAA cavity was installed in the vacuum chamber and connected to the rectangular waveguide via the mode converter.
We placed a Faraday cup in front of the mode converter to measure the dark current emitted from the beam hole in the coupler side.
We placed a pickup antenna composed of semi-rigid coax cable outside the copper endplate, along the beam axis.

The stored power $P_c$ coupled to the antenna was transmitted through the cable,
the attenuator (Lucas Weinschel Model 47), and a Keysight 8472B crystal diode, which generated a rectified voltage to be monitored by an oscilloscope (Iwatsu DS-5514).
The measurement using the signal generator indicated that
the attenuator damped the input RF power by $41~[{\rm dB}]$.

In the following, we describe the {\it in situ} measurements of the Q-value and the transmission coefficient representing the coupling of the antennas.
As shown in Figure~\ref{fig:cavity_rfsystem}~(f), the 4-port circulator is equipped with two vacuum windows, and the section between them is at atmospheric pressure and can be removed.
Thus, we could perform low-power measurements of the cavity through the waveguide from the position of the vacuum window.
The transmission coefficient $S_{12}$ from the waveguide to the antenna and coaxial cable (Huber \& Suhner Sucoflex 104) could be measured directly to obtain information on the coupling between the antenna and the cavity, including the transmission loss of the coaxial cable.
Neglecting the transmission loss by the rectangular waveguide,
we can obtain the ratio of the power observed by the antenna via the coaxial cable and that stored inside the cavity from the transmission coefficient $S_{12}$.

\begin{table}[htb]
  \begin{tabular}{cccccc} \hline
      Coating & $f_0~[{\rm MHz}]$ & $Q_L~(\times 10^4)$ & $\beta$ & $Q_0~(\times 10^4)$ & $S_{21}~[{\rm dB}]$\\\hline
      uncoated & $5736.81$ & $1.1$ & $1.5$ & $2.7$ & $32.7$\\
      DLC1 & $5739.72$ & $1.2$ & $1.1$ & $2.7$ & $26.7$\\
      DLC2 & $5729.65$ & $0.88$ & $2.1$ & $2.7$ & $33.1$\\ \hline
  \end{tabular}
    \caption{Results of the low-power test conducted using the network analyzer.}
  \label{tab:lowpower}
\end{table}

Table~\ref{tab:lowpower}
shows the results of the low-power measurement described above.
The rows represent the coating conditions,  i.e., without coating (uncoated) and with DLC coating with a thickness of $0.5~[{\rm \mu m}]$ (DLC1, DLC2).
Different pairs of dielectric cells were used in each condition, and the dimensional errors in the cells shifted the resonant frequency and coupling.
$f_0$ is the resonant frequency of the cavity.
$Q_L=Q_0/(1+\beta+\beta_{\rm a})$ is the loaded Q-value, where
the coupling between the cavity and the antenna satisfies $\beta_{\rm a}\ll1$ and
$\beta$ is the coupling through the iris of the coupler.

The stored power $P_c$ can be reconstructed via two independent methods using the measurements from the low-power test.

In the first method, the transient model is used to compute $P_c(t)$ from the input waveforms $P_{\rm in}(t)$, and the measured loaded Q value $Q_L$.
We assume that there is no loss due to time-varying multipactors in the cavity and then $Q_L$ is constant.
By using the relation of power conservation $dW_c/dt=P_{\rm in}-P_{\rm ref}-P_c^B$, $Q_0=\omega W_c/P_c^{\rm B}$, $P_{\rm in}=G_{\rm wg}V_{\rm in}^2$, $P_{\rm ref}=G_{\rm wg}(V_c - V_{\rm in})^2$, $P_c^B=(G_{\rm wg}/\beta)V_{\rm c}^2$, the cavity voltage $V_c$, and waveguide voltage follow
\begin{eqnarray}
  T_f\frac{dV_c}{dt} + V_c = \frac{2\beta}{1+\beta}V_{\rm in},\label{eq:dedt}
\end{eqnarray}
where $T_f = Q_L/(\pi f_0)$ and $G_{\rm wg}$ denotes the waveguide admittance.
By substituting the measured time variation of the incident power $P_{\rm in}(t)$ into Eq.~\eqref{eq:dedt}, we obtain the stored power $P_c^B(t)$.

The second method reconstructs the stored power $P_c^A$ using the transmission coefficients $S_{21}$, attenuator, and crystal diode.
Using the signal generator, we measured the relation between the input power and the output voltage, $f_{\rm dBm}(\ln V)$, for the crystal diode and RF power meters, where $V$ is the voltage measured by the oscilloscope.
Then, we can reconstruct the stored power $P_c^A$ as
\begin{eqnarray}
  P_c= 1~[{\rm mW}] \cdot10^{f_{\rm dBm}(\ln V)/10}
  \cdot 10^{-S_{21}~[{\rm dB}]/10} \cdot 10^{\Delta_{Att} /10},
\end{eqnarray}
where $\Delta_{Att}\sim 41$ is the power reduction measured by the attenuator.

From the simulation using {\tt Superfish}, the fraction of shunt impedance per unit length in the Q-value is obtained as
$Z/Q_0=4.25~[{\rm k\Omega/m}]$,
which is independent of the properties of the materials.
To obtain the shunt impedance per unit length $Z$, we multiplied the measured $Q_0$ by the fraction.
Combining the above quantities, we can obtain the peak accelerating voltage as
$E_{z,{\rm max}} = \sqrt{Z\cdot P_c/L}$, where we use a cavity length of $L=52.485~[{\rm mm}]$.

The repetition rate of the RF pulses is $50~[{\rm Hz}]$.
During the experiment, the RF frequency, magnitude of the incident power, and pulse length of the RF input from the klystron to the DAA cavity can be changed at any time.
We investigated the upper limit of the stored power by increasing the incident power from approximately 100~[W] to 100~[kW].
In this process, the RF frequency and pulse length were repeatedly adjusted to maximize the power waveform in the cavity measured by the antenna.
In the case of DLC1 and DLC2, the conditioning time to reach an accelerating field of 10 MV/m was approximately 6 h, which is shorter than the conditioning time of the normal-conducting cavity.

\begin{table}[htb]
  \begin{tabular}{cccccc} \hline
      Coating & $P_{c}^A~[{\rm kW}]$ & $P_c^A/P_c^B$ & $P_c^A/P_c^C$ & $E_{z}^A~[{\rm MV/m}]$& $N_{\rm shot}(\times 10^6)$ \\\hline
      uncoated & $0.89$ & $0.010$ & $0.013$ & $1.4$ & $1.5$ \\
      DLC1 & $56$ & $0.92$ & $0.74$ & $11$ & $1.5$ \\
      DLC2 & $48$ & $1.2$ & $1.0$ & $10$  & $3.8$ \\ \hline
  \end{tabular}

  \caption{Peak stored power in the cavity is calculated by two methods: (1) using signals from the pickup antenna and (2) using the pulse of the incident RF power. $P_c^C={\rm max}(P_{\rm in}(t)-P_{\rm ref}(t))$. $E_z^A$ is the peak accelerating gradient calculated by $P_c^A$. DLC1 and DLC2 denote the DLC-coated cells, with a thickness of $0.5~[{\rm \mu m}]$ on both sides.}
  \label{tab:tab_power}
\end{table}

\begin{figure*}[t]
\begin{tabular}{ccc}
  \begin{minipage}[t]{0.333\hsize}\centering\includegraphics[width=1.0\hsize]{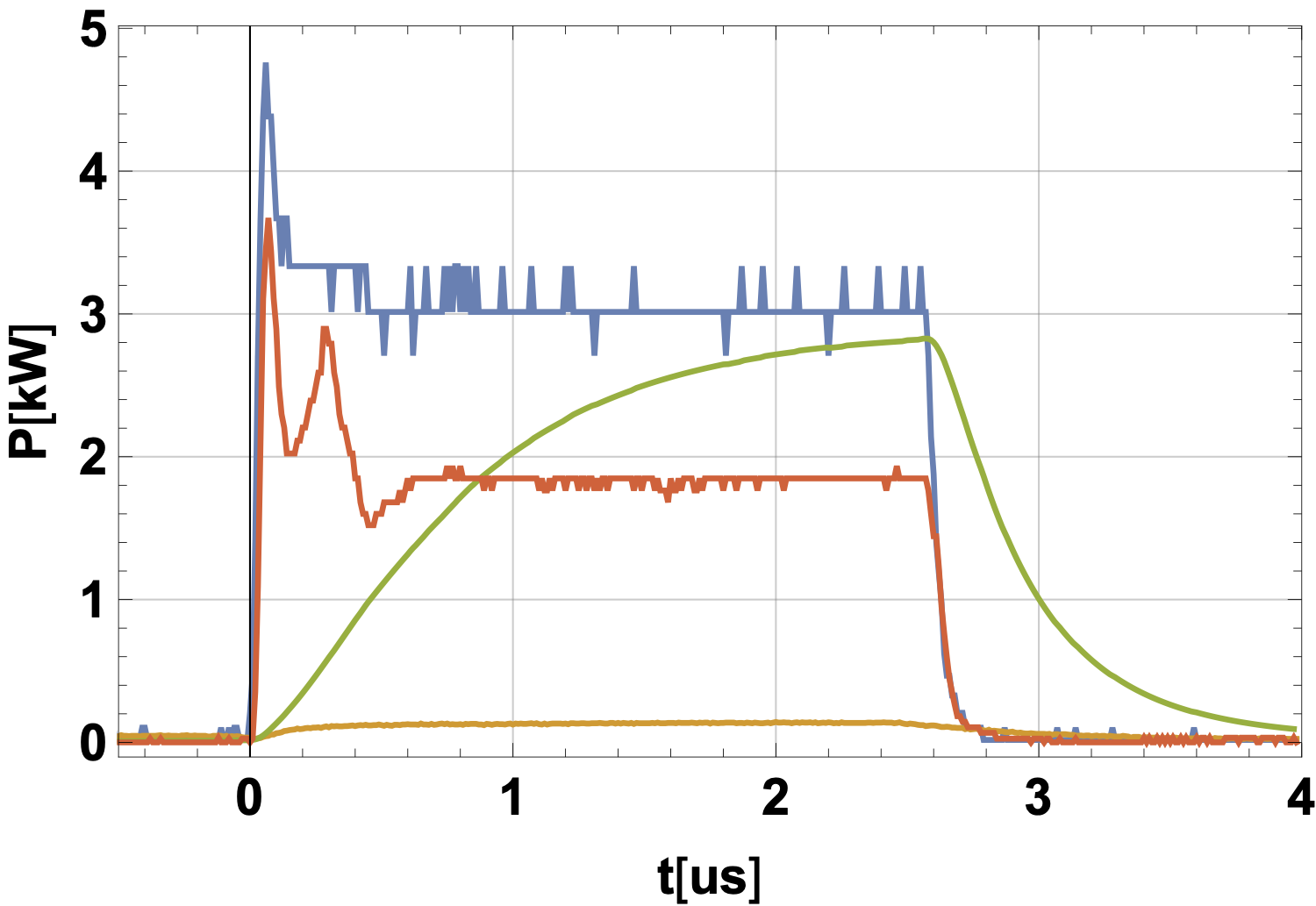}
    \caption{uncoated cells with multipactor}\label{fig:white}
  \end{minipage}
  \begin{minipage}[t]{0.333\hsize}\centering\includegraphics[width=1.0\hsize]{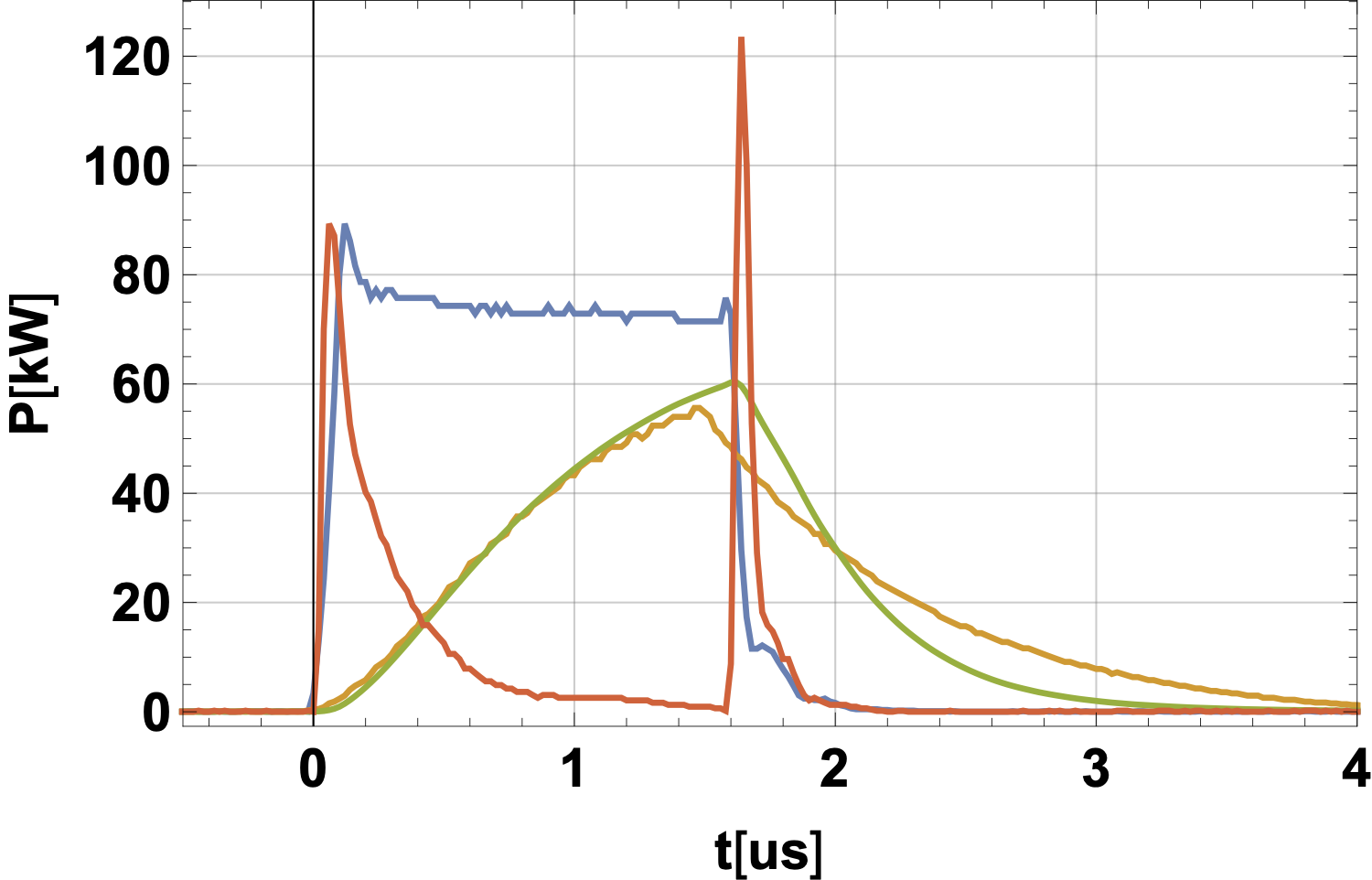}
    \caption{DLC1}\label{fig:dlc2}
  \end{minipage}
  \begin{minipage}[t]{0.333\hsize}\centering\includegraphics[width=1.0\hsize]{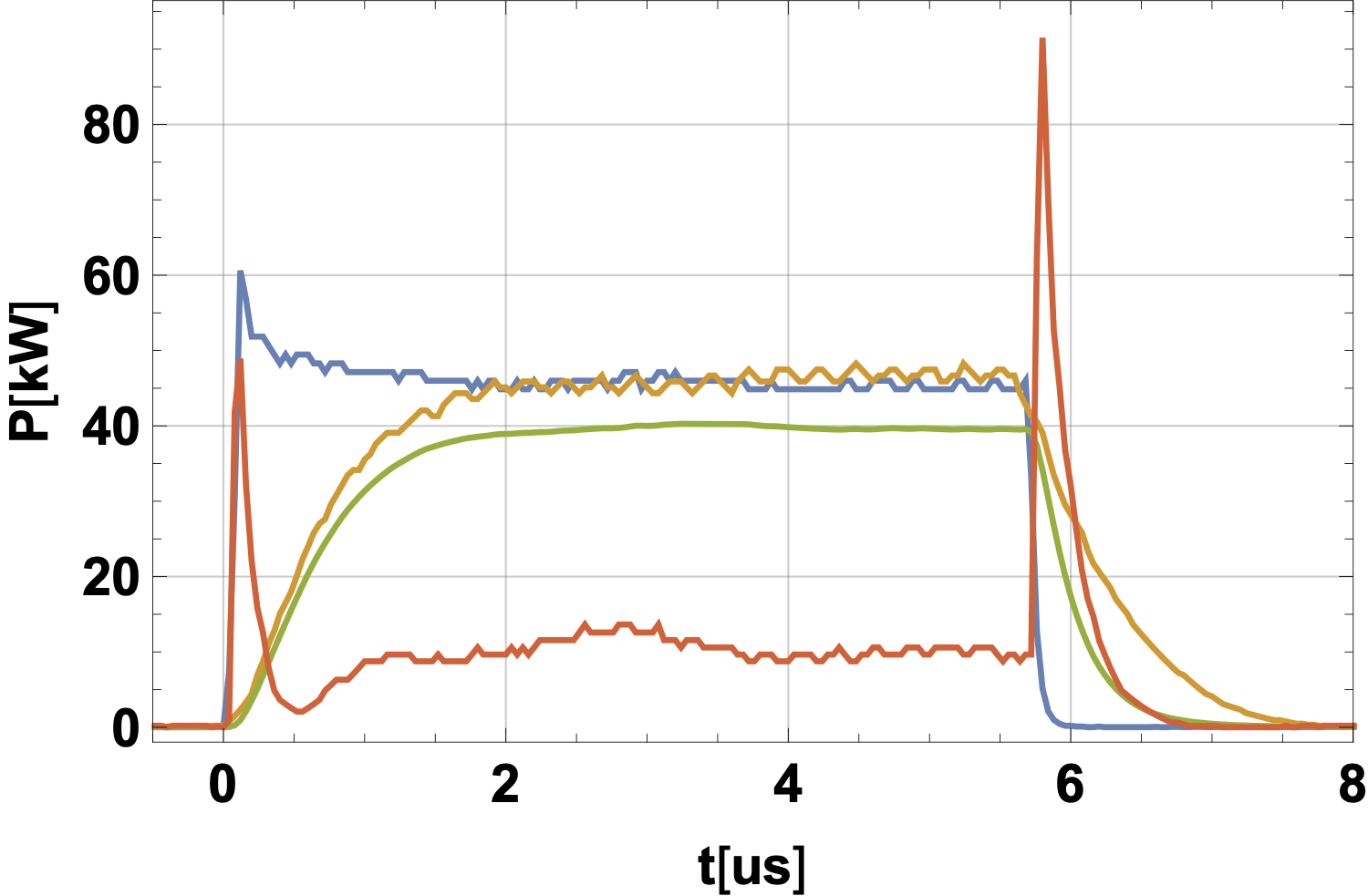}
    \caption{DLC2}\label{fig:dlc3}
  \end{minipage}
\end{tabular}
\caption{Waveforms of the measured incident power (blue), reflected power (red), and the stored power reconstructed by the antenna (yellow), $P_c^A$, and by the incident power (green), $P_c^B$.
For uncoated cells, typical waveforms with multipactors are shown.
For DLC1 and DLC2, the waveforms show stable stored power with an accelerating gradient exceeding 10 MV/m.}
\end{figure*}

In Figures~\ref{fig:white},~\ref{fig:dlc2}, and~\ref{fig:dlc3},
we show the waveforms of the measured incident power (blue), reflected power (red), and the stored power reconstructed by the antenna (yellow), $P_c^A$, and by the incident power (green), $P_c^B$.

As shown in Figures~\ref{fig:dlc2} and~\ref{fig:dlc3}, power equivalent to an accelerating electric field of over $10~[{\rm MV/m}]$ could be input when approximately $0.5~[{\rm \mu m}]$ of DLC was applied.
Table~\ref{tab:tab_power} shows each measured maximum stored power value and its conversion to the maximum accelerating electric field corresponding to these waveforms.
In the DLC1 test, when the accelerating electric field was increased further, a large discharge occurred and the pulses stopped coming in.
In the DLC2 test, the pulse length was extended to approximately $7~[{\rm \mu s}]$ while maintaining $10~[{\rm MV/m}]$, resulting in a waveform similar to that of DLC1.

\begin{figure}[t]
  \centering\includegraphics[width=0.9\hsize]{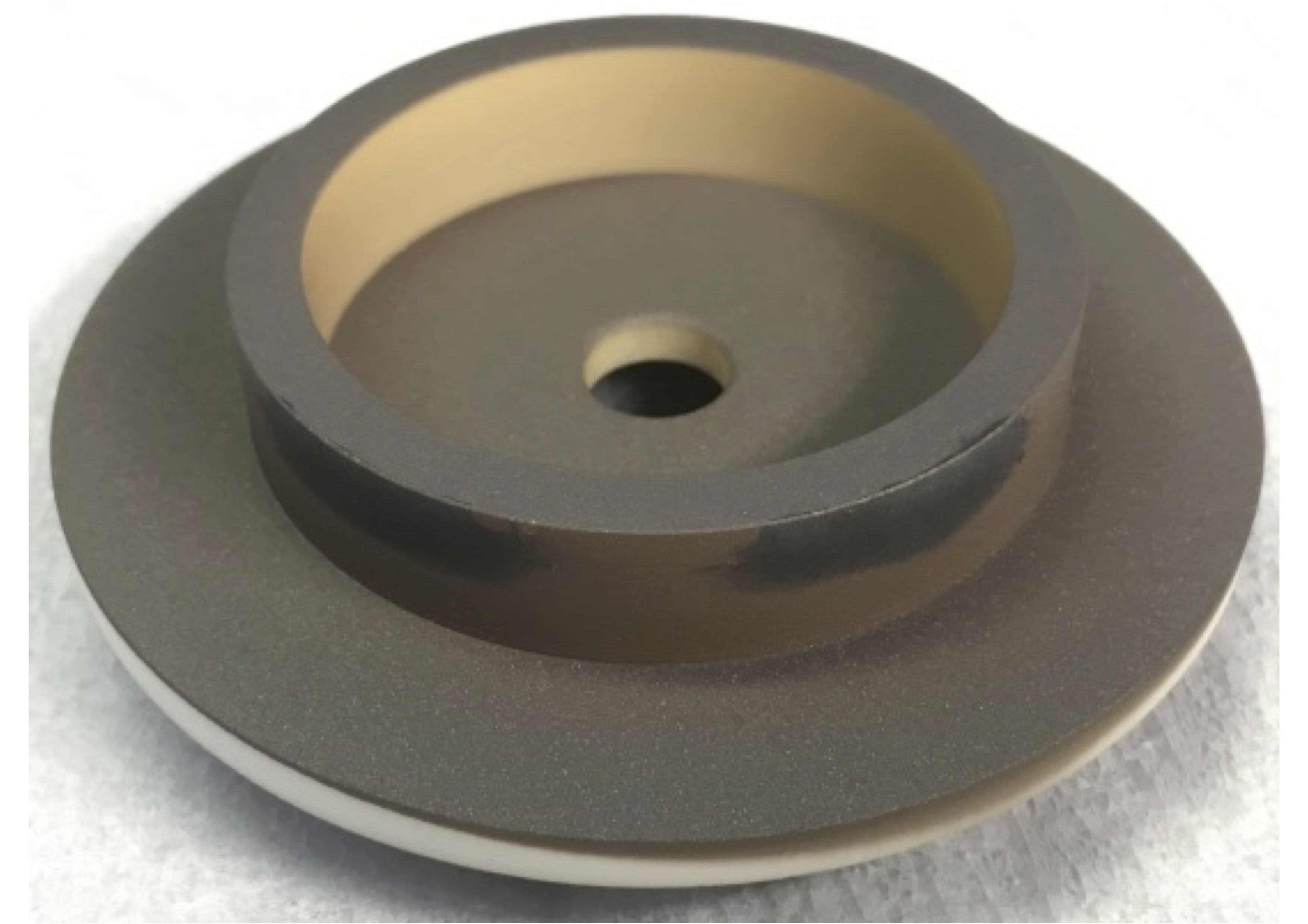}
  \caption{
  DLC-coated MgO cell after a breakdown in a high-power test. Black marks were observed at the corners of the cylindrical part of the cell. Black marks were also observed at the contact point between the dielectric cell and the adjacent dielectric cell and at the contact point between the dielectric and the copper end-plate.
  }
  \label{fig:trace}
\end{figure}

In an uncoated MgO cell, the stored power is saturated when the accelerating electric field is at the maximum electric field $E_{z,{\rm max}}\sim2~[{\rm MV/m}]$.
In this case, the saturation value of the stored power does not change even if $P_{\rm in}$ increases with the same pulse length $T_p$.
When the stored power is saturated, extending the pulse length $T_p$ can extend the region where the stored power waveform is flattened to $10~[{\rm MV/m}]$.
The reflected waveform is not consistent with the transient model; above $1~[{\rm MV/m}]$, the waveform is maintained at the same depth after a slight absorption.

When DLC coating was applied, the stored power increased in stages until a breakdown occurred in the cavity.
Once the breakdown occurred, the stored power waveform became unstable and did not improve with further conditioning.
After the test, some traces of electric discharge remained at the cell contact area and at the contact area between the cell and the metal end-plates, as shown in Figure.~\ref{fig:trace}.
Therefore, we consider that the microgap discharges at the cell junctions determine the maximum accelerating electric field in the DLC-coated DAA cavity.
In order to obtain a higher accelerating field, countermeasures against breakdowns at the cell junctions are required.
For example, methods of separating the dielectric cells from each other or separating the dielectric cells from the metal end-plates have been proposed in ~\cite{mori2019design,wei2020investigations}.

In conclusion,
By applying a DLC that lowers the SEY without reducing the Q-value, we have demonstrated that accelerating fields above $10~[{\rm MV/m}]$ can be applied for pulses as long as $6~[{\rm \mu s}]$, thus suppressing the multipactor that has limited the accelerating field of the DAA cavity.
We have also demonstrated that, owing to the very high resistance of the DLC coating, even a $0.5~[{\rm \mu m}]$-thick DLC coating on both sides of a dielectric cell does not reduce the Q-value ${\cal O}(10^5)$ of a five-cell DAA cavity.
One possible application of the DAA cavity is as a compact power source for X-ray and RI production, as DAA cavities have a power efficiency one order of magnitude higher than that of a normal-conducting cavity.
In the future, if the anti-breakdown treatment is successful and the accelerating field can be further extended to tens of MV/m, a linear accelerator facility such as an FEL may be able to operate at room temperature without RF pulse compression by the SLED cavity.

\section*{Acknowledgements}
We would like to acknowledge Hiroyasu Ego for helpful discussions.
This work was supported by JSPS KAKENHI Grant-in-Aid
for Scientific Research (A) (No.~16H02134),
JSPS KAKENHI Grant Number 19K20609,
and Mitsubishi Heavy Industries Machinery Systems, Ltd.

The data that support the findings of this study are available from the corresponding author
upon reasonable request.

\bibliography{ref}

\begin{thebibliography}{19}%
\makeatletter
\providecommand \@ifxundefined [1]{%
 \@ifx{#1\undefined}
}%
\providecommand \@ifnum [1]{%
 \ifnum #1\expandafter \@firstoftwo
 \else \expandafter \@secondoftwo
 \fi
}%
\providecommand \@ifx [1]{%
 \ifx #1\expandafter \@firstoftwo
 \else \expandafter \@secondoftwo
 \fi
}%
\providecommand \natexlab [1]{#1}%
\providecommand \enquote  [1]{``#1''}%
\providecommand \bibnamefont  [1]{#1}%
\providecommand \bibfnamefont [1]{#1}%
\providecommand \citenamefont [1]{#1}%
\providecommand \href@noop [0]{\@secondoftwo}%
\providecommand \href [0]{\begingroup \@sanitize@url \@href}%
\providecommand \@href[1]{\@@startlink{#1}\@@href}%
\providecommand \@@href[1]{\endgroup#1\@@endlink}%
\providecommand \@sanitize@url [0]{\catcode `\\12\catcode `\$12\catcode
  `\&12\catcode `\#12\catcode `\^12\catcode `\_12\catcode `\%12\relax}%
\providecommand \@@startlink[1]{}%
\providecommand \@@endlink[0]{}%
\providecommand \url  [0]{\begingroup\@sanitize@url \@url }%
\providecommand \@url [1]{\endgroup\@href {#1}{\urlprefix }}%
\providecommand \urlprefix  [0]{URL }%
\providecommand \Eprint [0]{\href }%
\providecommand \doibase [0]{https://doi.org/}%
\providecommand \selectlanguage [0]{\@gobble}%
\providecommand \bibinfo  [0]{\@secondoftwo}%
\providecommand \bibfield  [0]{\@secondoftwo}%
\providecommand \translation [1]{[#1]}%
\providecommand \BibitemOpen [0]{}%
\providecommand \bibitemStop [0]{}%
\providecommand \bibitemNoStop [0]{.\EOS\space}%
\providecommand \EOS [0]{\spacefactor3000\relax}%
\providecommand \BibitemShut  [1]{\csname bibitem#1\endcsname}%
\let\auto@bib@innerbib\@empty
%</preamble>
\bibitem [{\citenamefont {Nyaiesh}\ \emph {et~al.}(1986)\citenamefont
  {Nyaiesh}, \citenamefont {Garwin}, \citenamefont {King},\ and\ \citenamefont
  {Kirby}}]{nyaiesh1986properties}%
  \BibitemOpen
  \bibfield  {author} {\bibinfo {author} {\bibfnamefont {A.}~\bibnamefont
  {Nyaiesh}}, \bibinfo {author} {\bibfnamefont {E.}~\bibnamefont {Garwin}},
  \bibinfo {author} {\bibfnamefont {F.}~\bibnamefont {King}},\ and\ \bibinfo
  {author} {\bibfnamefont {R.}~\bibnamefont {Kirby}},\ }\href@noop {}
  {\bibfield  {journal} {\bibinfo  {journal} {Journal of Vacuum Science \&
  Technology A: Vacuum, Surfaces, and Films}\ }\textbf {\bibinfo {volume}
  {4}},\ \bibinfo {pages} {2356} (\bibinfo {year} {1986})}\BibitemShut
  {NoStop}%
\bibitem [{\citenamefont {Kennedy}\ \emph {et~al.}(1997)\citenamefont
  {Kennedy}, \citenamefont {Harteneck}, \citenamefont {Millos}, \citenamefont
  {Benapfl}, \citenamefont {King},\ and\ \citenamefont
  {Kirby}}]{kennedy1997tin}%
  \BibitemOpen
  \bibfield  {author} {\bibinfo {author} {\bibfnamefont {K.}~\bibnamefont
  {Kennedy}}, \bibinfo {author} {\bibfnamefont {B.}~\bibnamefont {Harteneck}},
  \bibinfo {author} {\bibfnamefont {G.}~\bibnamefont {Millos}}, \bibinfo
  {author} {\bibfnamefont {M.}~\bibnamefont {Benapfl}}, \bibinfo {author}
  {\bibfnamefont {F.}~\bibnamefont {King}},\ and\ \bibinfo {author}
  {\bibfnamefont {R.}~\bibnamefont {Kirby}},\ }in\ \href@noop {} {\emph
  {\bibinfo {booktitle} {Proceedings of the 1997 Particle Accelerator
  Conference (Cat. No. 97CH36167)}}},\ Vol.~\bibinfo {volume} {3}\ (\bibinfo
  {organization} {IEEE},\ \bibinfo {year} {1997})\ pp.\ \bibinfo {pages}
  {3568--3570}\BibitemShut {NoStop}%
\bibitem [{\citenamefont {Vallgren}\ \emph {et~al.}(2011)\citenamefont
  {Vallgren}, \citenamefont {Arduini}, \citenamefont {Bauche}, \citenamefont
  {Calatroni}, \citenamefont {Chiggiato}, \citenamefont {Cornelis},
  \citenamefont {Pinto}, \citenamefont {Henrist}, \citenamefont {M{\'e}tral},
  \citenamefont {Neupert} \emph {et~al.}}]{vallgren2011amorphous}%
  \BibitemOpen
  \bibfield  {author} {\bibinfo {author} {\bibfnamefont {C.~Y.}\ \bibnamefont
  {Vallgren}}, \bibinfo {author} {\bibfnamefont {G.}~\bibnamefont {Arduini}},
  \bibinfo {author} {\bibfnamefont {J.}~\bibnamefont {Bauche}}, \bibinfo
  {author} {\bibfnamefont {S.}~\bibnamefont {Calatroni}}, \bibinfo {author}
  {\bibfnamefont {P.}~\bibnamefont {Chiggiato}}, \bibinfo {author}
  {\bibfnamefont {K.}~\bibnamefont {Cornelis}}, \bibinfo {author}
  {\bibfnamefont {P.~C.}\ \bibnamefont {Pinto}}, \bibinfo {author}
  {\bibfnamefont {B.}~\bibnamefont {Henrist}}, \bibinfo {author} {\bibfnamefont
  {E.}~\bibnamefont {M{\'e}tral}}, \bibinfo {author} {\bibfnamefont
  {H.}~\bibnamefont {Neupert}}, \emph {et~al.},\ }\href@noop {} {\bibfield
  {journal} {\bibinfo  {journal} {Physical Review Special Topics-Accelerators
  and Beams}\ }\textbf {\bibinfo {volume} {14}},\ \bibinfo {pages} {071001}
  (\bibinfo {year} {2011})}\BibitemShut {NoStop}%
\bibitem [{\citenamefont {Vallgren}\ \emph {et~al.}(2010)\citenamefont
  {Vallgren}, \citenamefont {Arduini}, \citenamefont {Bauche}, \citenamefont
  {Calatroni}, \citenamefont {Chiggiato}, \citenamefont {Cornelis},
  \citenamefont {Pinto}, \citenamefont {M{\'e}tral}, \citenamefont {Rumolo},
  \citenamefont {Shaposhnikova} \emph {et~al.}}]{vallgren2010amorphous}%
  \BibitemOpen
  \bibfield  {author} {\bibinfo {author} {\bibfnamefont {C.~Y.}\ \bibnamefont
  {Vallgren}}, \bibinfo {author} {\bibfnamefont {G.}~\bibnamefont {Arduini}},
  \bibinfo {author} {\bibfnamefont {J.}~\bibnamefont {Bauche}}, \bibinfo
  {author} {\bibfnamefont {S.}~\bibnamefont {Calatroni}}, \bibinfo {author}
  {\bibfnamefont {P.}~\bibnamefont {Chiggiato}}, \bibinfo {author}
  {\bibfnamefont {K.}~\bibnamefont {Cornelis}}, \bibinfo {author}
  {\bibfnamefont {P.~C.}\ \bibnamefont {Pinto}}, \bibinfo {author}
  {\bibfnamefont {E.}~\bibnamefont {M{\'e}tral}}, \bibinfo {author}
  {\bibfnamefont {G.}~\bibnamefont {Rumolo}}, \bibinfo {author} {\bibfnamefont
  {E.}~\bibnamefont {Shaposhnikova}}, \emph {et~al.},\ }in\ \href@noop {}
  {\emph {\bibinfo {booktitle} {1st International Particle Accelerator
  Conference: IPAC}}},\ Vol.~\bibinfo {volume} {10}\ (\bibinfo {year}
  {2010})\BibitemShut {NoStop}%
\bibitem [{\citenamefont {Shaposhnikova}\ \emph {et~al.}(2009)\citenamefont
  {Shaposhnikova}, \citenamefont {Rumolo}, \citenamefont {Cornelis},
  \citenamefont {Axensalva}, \citenamefont {Jim{\'e}nez}, \citenamefont
  {Taborelli}, \citenamefont {Chiggiato}, \citenamefont {Calatroni},
  \citenamefont {Arduini}, \citenamefont {Yin~Vallgren} \emph
  {et~al.}}]{shaposhnikova2009experimental}%
  \BibitemOpen
  \bibfield  {author} {\bibinfo {author} {\bibfnamefont {E.}~\bibnamefont
  {Shaposhnikova}}, \bibinfo {author} {\bibfnamefont {G.}~\bibnamefont
  {Rumolo}}, \bibinfo {author} {\bibfnamefont {K.}~\bibnamefont {Cornelis}},
  \bibinfo {author} {\bibfnamefont {J.}~\bibnamefont {Axensalva}}, \bibinfo
  {author} {\bibfnamefont {J.}~\bibnamefont {Jim{\'e}nez}}, \bibinfo {author}
  {\bibfnamefont {M.}~\bibnamefont {Taborelli}}, \bibinfo {author}
  {\bibfnamefont {P.}~\bibnamefont {Chiggiato}}, \bibinfo {author}
  {\bibfnamefont {S.}~\bibnamefont {Calatroni}}, \bibinfo {author}
  {\bibfnamefont {G.}~\bibnamefont {Arduini}}, \bibinfo {author} {\bibfnamefont
  {C.}~\bibnamefont {Yin~Vallgren}}, \emph {et~al.},\ }\href@noop {} {\emph
  {\bibinfo {title} {Experimental studies of carbon coatings as possible means
  of suppressing beam induced electron multipacting in the CERN SPS}}},\
  \bibinfo {type} {Tech. Rep.}\ (\bibinfo {year} {2009})\BibitemShut {NoStop}%
\bibitem [{\citenamefont {Baglin}\ \emph {et~al.}(1998)\citenamefont {Baglin},
  \citenamefont {Collins},\ and\ \citenamefont
  {Gr{\"o}bner}}]{baglin1998photoelectron}%
  \BibitemOpen
  \bibfield  {author} {\bibinfo {author} {\bibfnamefont {V.}~\bibnamefont
  {Baglin}}, \bibinfo {author} {\bibfnamefont {I.}~\bibnamefont {Collins}},\
  and\ \bibinfo {author} {\bibfnamefont {O.}~\bibnamefont {Gr{\"o}bner}},\
  }\href@noop {} {\emph {\bibinfo {title} {Photoelectron yield and photon
  reflectivity from candidate LHC vacuum chamber materials with implications to
  the vacuum chamber design}}},\ \bibinfo {type} {Tech. Rep.}\ (\bibinfo {year}
  {1998})\BibitemShut {NoStop}%
\bibitem [{\citenamefont {Kulikov}\ \emph {et~al.}(2001)\citenamefont
  {Kulikov}, \citenamefont {Fisher}, \citenamefont {Heifets}, \citenamefont
  {Seeman}, \citenamefont {Sullivan}, \citenamefont {Wienands},\ and\
  \citenamefont {Kozanecki}}]{kulikov2001electron}%
  \BibitemOpen
  \bibfield  {author} {\bibinfo {author} {\bibfnamefont {A.}~\bibnamefont
  {Kulikov}}, \bibinfo {author} {\bibfnamefont {A.}~\bibnamefont {Fisher}},
  \bibinfo {author} {\bibfnamefont {S.}~\bibnamefont {Heifets}}, \bibinfo
  {author} {\bibfnamefont {J.}~\bibnamefont {Seeman}}, \bibinfo {author}
  {\bibfnamefont {M.}~\bibnamefont {Sullivan}}, \bibinfo {author}
  {\bibfnamefont {U.}~\bibnamefont {Wienands}},\ and\ \bibinfo {author}
  {\bibfnamefont {W.}~\bibnamefont {Kozanecki}},\ }in\ \href@noop {} {\emph
  {\bibinfo {booktitle} {PACS2001. Proceedings of the 2001 Particle Accelerator
  Conference (Cat. No. 01CH37268)}}},\ Vol.~\bibinfo {volume} {3}\ (\bibinfo
  {organization} {IEEE},\ \bibinfo {year} {2001})\ pp.\ \bibinfo {pages}
  {1903--1905}\BibitemShut {NoStop}%
\bibitem [{\citenamefont {Jing}\ \emph {et~al.}(2016)\citenamefont {Jing},
  \citenamefont {Gold}, \citenamefont {Fischer},\ and\ \citenamefont
  {Gai}}]{jing2016complete}%
  \BibitemOpen
  \bibfield  {author} {\bibinfo {author} {\bibfnamefont {C.}~\bibnamefont
  {Jing}}, \bibinfo {author} {\bibfnamefont {S.}~\bibnamefont {Gold}}, \bibinfo
  {author} {\bibfnamefont {R.}~\bibnamefont {Fischer}},\ and\ \bibinfo {author}
  {\bibfnamefont {W.}~\bibnamefont {Gai}},\ }\href@noop {} {\bibfield
  {journal} {\bibinfo  {journal} {Applied Physics Letters}\ }\textbf {\bibinfo
  {volume} {108}},\ \bibinfo {pages} {193501} (\bibinfo {year}
  {2016})}\BibitemShut {NoStop}%
\bibitem [{\citenamefont {Satoh}\ \emph {et~al.}(2016)\citenamefont {Satoh},
  \citenamefont {Yoshida},\ and\ \citenamefont
  {Hayashizaki}}]{satoh2016dielectric}%
  \BibitemOpen
  \bibfield  {author} {\bibinfo {author} {\bibfnamefont {D.}~\bibnamefont
  {Satoh}}, \bibinfo {author} {\bibfnamefont {M.}~\bibnamefont {Yoshida}},\
  and\ \bibinfo {author} {\bibfnamefont {N.}~\bibnamefont {Hayashizaki}},\
  }\href@noop {} {\bibfield  {journal} {\bibinfo  {journal} {Physical Review
  Accelerators and Beams}\ }\textbf {\bibinfo {volume} {19}},\ \bibinfo {pages}
  {011302} (\bibinfo {year} {2016})}\BibitemShut {NoStop}%
\bibitem [{\citenamefont {Satoh}\ \emph {et~al.}(2017)\citenamefont {Satoh},
  \citenamefont {Yoshida},\ and\ \citenamefont
  {Hayashizaki}}]{satoh2017fabrication}%
  \BibitemOpen
  \bibfield  {author} {\bibinfo {author} {\bibfnamefont {D.}~\bibnamefont
  {Satoh}}, \bibinfo {author} {\bibfnamefont {M.}~\bibnamefont {Yoshida}},\
  and\ \bibinfo {author} {\bibfnamefont {N.}~\bibnamefont {Hayashizaki}},\
  }\href@noop {} {\bibfield  {journal} {\bibinfo  {journal} {Physical Review
  Accelerators and Beams}\ }\textbf {\bibinfo {volume} {20}},\ \bibinfo {pages}
  {091302} (\bibinfo {year} {2017})}\BibitemShut {NoStop}%
\bibitem [{\citenamefont {Power}\ \emph {et~al.}(2004)\citenamefont {Power},
  \citenamefont {Gai}, \citenamefont {Gold}, \citenamefont {Kinkead},
  \citenamefont {Konecny}, \citenamefont {Jing}, \citenamefont {Liu},\ and\
  \citenamefont {Yusof}}]{power2004observation}%
  \BibitemOpen
  \bibfield  {author} {\bibinfo {author} {\bibfnamefont {J.}~\bibnamefont
  {Power}}, \bibinfo {author} {\bibfnamefont {W.}~\bibnamefont {Gai}}, \bibinfo
  {author} {\bibfnamefont {S.}~\bibnamefont {Gold}}, \bibinfo {author}
  {\bibfnamefont {A.}~\bibnamefont {Kinkead}}, \bibinfo {author} {\bibfnamefont
  {R.}~\bibnamefont {Konecny}}, \bibinfo {author} {\bibfnamefont
  {C.}~\bibnamefont {Jing}}, \bibinfo {author} {\bibfnamefont {W.}~\bibnamefont
  {Liu}},\ and\ \bibinfo {author} {\bibfnamefont {Z.}~\bibnamefont {Yusof}},\
  }\href@noop {} {\bibfield  {journal} {\bibinfo  {journal} {Physical review
  letters}\ }\textbf {\bibinfo {volume} {92}},\ \bibinfo {pages} {164801}
  (\bibinfo {year} {2004})}\BibitemShut {NoStop}%
\bibitem [{\citenamefont {Zhang}\ \emph {et~al.}(2016)\citenamefont {Zhang},
  \citenamefont {Munroe}, \citenamefont {Xu}, \citenamefont {Shapiro},\ and\
  \citenamefont {Temkin}}]{zhang2016high}%
  \BibitemOpen
  \bibfield  {author} {\bibinfo {author} {\bibfnamefont {J.}~\bibnamefont
  {Zhang}}, \bibinfo {author} {\bibfnamefont {B.~J.}\ \bibnamefont {Munroe}},
  \bibinfo {author} {\bibfnamefont {H.}~\bibnamefont {Xu}}, \bibinfo {author}
  {\bibfnamefont {M.~A.}\ \bibnamefont {Shapiro}},\ and\ \bibinfo {author}
  {\bibfnamefont {R.~J.}\ \bibnamefont {Temkin}},\ }\href@noop {} {\bibfield
  {journal} {\bibinfo  {journal} {Physical Review Accelerators and Beams}\
  }\textbf {\bibinfo {volume} {19}},\ \bibinfo {pages} {081304} (\bibinfo
  {year} {2016})}\BibitemShut {NoStop}%
\bibitem [{\citenamefont {Jing}\ \emph {et~al.}(2010)\citenamefont {Jing},
  \citenamefont {Gai}, \citenamefont {Power}, \citenamefont {Konecny},
  \citenamefont {Liu}, \citenamefont {Gold}, \citenamefont {Kinkead},
  \citenamefont {Tantawi}, \citenamefont {Dolgashev},\ and\ \citenamefont
  {Kanareykin}}]{jing2010progress}%
  \BibitemOpen
  \bibfield  {author} {\bibinfo {author} {\bibfnamefont {C.}~\bibnamefont
  {Jing}}, \bibinfo {author} {\bibfnamefont {W.}~\bibnamefont {Gai}}, \bibinfo
  {author} {\bibfnamefont {J.~G.}\ \bibnamefont {Power}}, \bibinfo {author}
  {\bibfnamefont {R.}~\bibnamefont {Konecny}}, \bibinfo {author} {\bibfnamefont
  {W.}~\bibnamefont {Liu}}, \bibinfo {author} {\bibfnamefont {S.~H.}\
  \bibnamefont {Gold}}, \bibinfo {author} {\bibfnamefont {A.~K.}\ \bibnamefont
  {Kinkead}}, \bibinfo {author} {\bibfnamefont {S.~G.}\ \bibnamefont
  {Tantawi}}, \bibinfo {author} {\bibfnamefont {V.}~\bibnamefont {Dolgashev}},\
  and\ \bibinfo {author} {\bibfnamefont {A.}~\bibnamefont {Kanareykin}},\
  }\href@noop {} {\bibfield  {journal} {\bibinfo  {journal} {IEEE transactions
  on plasma science}\ }\textbf {\bibinfo {volume} {38}},\ \bibinfo {pages}
  {1354} (\bibinfo {year} {2010})}\BibitemShut {NoStop}%
\bibitem [{\citenamefont {Xu}\ \emph {et~al.}(2019)\citenamefont {Xu},
  \citenamefont {Shapiro},\ and\ \citenamefont {Temkin}}]{xu2019measurement}%
  \BibitemOpen
  \bibfield  {author} {\bibinfo {author} {\bibfnamefont {H.}~\bibnamefont
  {Xu}}, \bibinfo {author} {\bibfnamefont {M.}~\bibnamefont {Shapiro}},\ and\
  \bibinfo {author} {\bibfnamefont {R.~J.}\ \bibnamefont {Temkin}},\
  }\href@noop {} {\bibfield  {journal} {\bibinfo  {journal} {Physical Review
  Accelerators and Beams}\ }\textbf {\bibinfo {volume} {22}},\ \bibinfo {pages}
  {021002} (\bibinfo {year} {2019})}\BibitemShut {NoStop}%
\bibitem [{\citenamefont {Zhang}\ \emph {et~al.}(2019)\citenamefont {Zhang},
  \citenamefont {Wang}, \citenamefont {Ge}, \citenamefont {Zhang},
  \citenamefont {Wei}, \citenamefont {Wang}, \citenamefont {Zhu}, \citenamefont
  {Shao}, \citenamefont {Li},\ and\ \citenamefont {Wang}}]{zhang2019research}%
  \BibitemOpen
  \bibfield  {author} {\bibinfo {author} {\bibfnamefont {Y.}~\bibnamefont
  {Zhang}}, \bibinfo {author} {\bibfnamefont {Y.}~\bibnamefont {Wang}},
  \bibinfo {author} {\bibfnamefont {X.}~\bibnamefont {Ge}}, \bibinfo {author}
  {\bibfnamefont {B.}~\bibnamefont {Zhang}}, \bibinfo {author} {\bibfnamefont
  {W.}~\bibnamefont {Wei}}, \bibinfo {author} {\bibfnamefont {S.}~\bibnamefont
  {Wang}}, \bibinfo {author} {\bibfnamefont {B.}~\bibnamefont {Zhu}}, \bibinfo
  {author} {\bibfnamefont {J.}~\bibnamefont {Shao}}, \bibinfo {author}
  {\bibfnamefont {W.}~\bibnamefont {Li}},\ and\ \bibinfo {author}
  {\bibfnamefont {Y.}~\bibnamefont {Wang}},\ }in\ \href@noop {} {\emph
  {\bibinfo {booktitle} {Journal of Physics: Conference Series}}},\ Vol.\
  \bibinfo {volume} {1350}\ (\bibinfo {organization} {IOP Publishing},\
  \bibinfo {year} {2019})\ p.\ \bibinfo {pages} {012177}\BibitemShut {NoStop}%
\bibitem [{\citenamefont {Yamamoto}\ \emph {et~al.}(2019)\citenamefont
  {Yamamoto}, \citenamefont {Cenni}, \citenamefont {Four}, \citenamefont
  {Kako}, \citenamefont {Michizono},\ and\ \citenamefont
  {Okii}}]{yamamoto2019recent}%
  \BibitemOpen
  \bibfield  {author} {\bibinfo {author} {\bibfnamefont {Y.}~\bibnamefont
  {Yamamoto}}, \bibinfo {author} {\bibfnamefont {E.}~\bibnamefont {Cenni}},
  \bibinfo {author} {\bibfnamefont {A.}~\bibnamefont {Four}}, \bibinfo {author}
  {\bibfnamefont {E.}~\bibnamefont {Kako}}, \bibinfo {author} {\bibfnamefont
  {S.}~\bibnamefont {Michizono}},\ and\ \bibinfo {author} {\bibfnamefont
  {Y.}~\bibnamefont {Okii}},\ }in\ \href@noop {} {\emph {\bibinfo {booktitle}
  {29$\backslash$textsuperscript $\{$th$\}$ Linear Accelerator Conf.(LINAC'18),
  Beijing, China, 16-21 September 2018}}}\ (\bibinfo {organization} {JACOW
  Publishing, Geneva, Switzerland},\ \bibinfo {year} {2019})\ pp.\ \bibinfo
  {pages} {905--907}\BibitemShut {NoStop}%
\bibitem [{\citenamefont {Meyerson}\ and\ \citenamefont
  {Smith}(1980)}]{meyerson1980electrical}%
  \BibitemOpen
  \bibfield  {author} {\bibinfo {author} {\bibfnamefont {B.}~\bibnamefont
  {Meyerson}}\ and\ \bibinfo {author} {\bibfnamefont {F.}~\bibnamefont
  {Smith}},\ }\href@noop {} {\bibfield  {journal} {\bibinfo  {journal} {Journal
  of Non-Crystalline Solids}\ }\textbf {\bibinfo {volume} {35}},\ \bibinfo
  {pages} {435} (\bibinfo {year} {1980})}\BibitemShut {NoStop}%
\bibitem [{\citenamefont {Mori}\ \emph {et~al.}(2019)\citenamefont {Mori},
  \citenamefont {Yoshida},\ and\ \citenamefont {Satoh}}]{mori2019design}%
  \BibitemOpen
  \bibfield  {author} {\bibinfo {author} {\bibfnamefont {S.}~\bibnamefont
  {Mori}}, \bibinfo {author} {\bibfnamefont {M.}~\bibnamefont {Yoshida}},\ and\
  \bibinfo {author} {\bibfnamefont {J.~D.}\ \bibnamefont {Satoh}},\ }in\
  \href@noop {} {\emph {\bibinfo {booktitle} {10th Int. Partile Accelerator
  Conf.(IPAC'19), Melbourne, Australia, 19-24 May 2019}}}\ (\bibinfo
  {organization} {JACOW Publishing, Geneva, Switzerland},\ \bibinfo {year}
  {2019})\ pp.\ \bibinfo {pages} {1179--1181}\BibitemShut {NoStop}%
\bibitem [{\citenamefont {Wei}\ and\ \citenamefont
  {Grudiev}(2020)}]{wei2020investigations}%
  \BibitemOpen
  \bibfield  {author} {\bibinfo {author} {\bibfnamefont {Y.}~\bibnamefont
  {Wei}}\ and\ \bibinfo {author} {\bibfnamefont {A.}~\bibnamefont {Grudiev}},\
  }\href@noop {} {\bibfield  {journal} {\bibinfo  {journal} {arXiv preprint
  arXiv:2006.07276}\ } (\bibinfo {year} {2020})}\BibitemShut {NoStop}%
\end{thebibliography}%
\bibliographystyle{apsrev4-2}
\end{document}